\documentclass[11pt]{article}
\usepackage{amssymb}
\usepackage{amsmath}
\usepackage[english]{babel}
\usepackage{graphicx}

\def\Journal#1#2#3#4{{#1} {\bf #2}, #3 (#4)}

\def\BWP{\em Bled Workshops in Physics}

\def\({\left(}
\def\){\right)}

\def\beq{\begin{equation}}
\def\eeq{\end{equation}}
\def\bea{\begin{eqnarray}}
\def\eea{\end{eqnarray}}

\begin{document}
\begin{center}
        \large \textbf{VIA Discussions at XIV Bled Workshop}
    \end{center}
\begin{center}
   Maxim Yu. Khlopov$^{1,2,3}$

    \emph{$^{1}$Moscow Engineering Physics Institute (National Nuclear Research University), 115409 Moscow, Russia \\
    $^{2}$ Centre for Cosmoparticle Physics "Cosmion" 125047 Moscow, Russia \\
$^{3}$ APC laboratory 10, rue Alice Domon et L\'eonie Duquet \\75205
Paris Cedex 13, France}

    \end{center}

\medskip
\begin{abstract}

Virtual Institute of Astroparticle Physics (VIA), integrated in the
structure of Laboratory of AstroParticle physics and Cosmology (APC)
is evolved in a unique multi-functional complex of $e-science$ and $e-learning$, supporting various
forms of collaborative scientific work as well as programs of education at
distance. The activity of VIA takes place on its website and
includes regular videoconferences with systematic basic courses and
lectures on various issues of astroparticle physics, regular online
transmission of APC Colloquiums, participation at distance in
various scientific meetings and conferences, library of their
records and presentations, a multilingual Forum. VIA virtual rooms
are open for meetings of scientific groups and for individual work
of supervisors with their students. The format of a VIA
videoconferences was effectively used in the program of XIV Bled
Workshop to provide a world-wide participation at distance in
discussion of the open questions of physics beyond the standard
model. The VIA system has demonstrated its high quality and stability even for minimal equipment (laptop with microphone and webcam and WiFi Internet connection).

\end{abstract}

\section{Introduction}
Studies in astroparticle physics link astrophysics, cosmology,
particle and nuclear physics and involve hundreds of scientific
groups linked by regional networks (like ASPERA/ApPEC \cite{aspera})
and national centers. The exciting progress in these studies will
have impact on the fundamental knowledge on the structure of
microworld and Universe and on the basic, still unknown, physical
laws of Nature (see e.g. \cite{book,newbook} for review).

In the proposal \cite{Khlopov:2008vd} it was suggested to organize a
Virtual Institute of Astroparticle Physics (VIA), which can play the
role of an unifying and coordinating structure for astroparticle
physics. Starting from the January of 2008 the activity of the
Institute takes place on its website \cite{VIA} in a form of regular
weekly videoconferences with VIA lectures, covering all the
theoretical and experimental activities in astroparticle physics and
related topics. The library of records of these lectures, talks and
their presentations is now accomplished by multi-lingual Forum. In
2008 VIA complex was effectively used for the first time for
participation at distance in XI Bled Workshop \cite{archiVIA}. Since
then VIA videoconferences became a natural part of Bled Workshops'
programs, opening the virtual room of discussions to the world-wide
audience. Its progress was presented in \cite{BledVIA9,BledVIA10}. Here the
current state-of-art of VIA complex, integrated since the end of
2009 in the structure of APC Laboratory, is presented in order to
clarify the way in which VIA discussion of open questions beyond the
standard model took place in the framework of XIV Bled Workshop.
\section{The current structure of VIA complex}
\subsection{The forms of VIA activity}
\begin{figure}
    \begin{center}
        \includegraphics[scale=0.3]{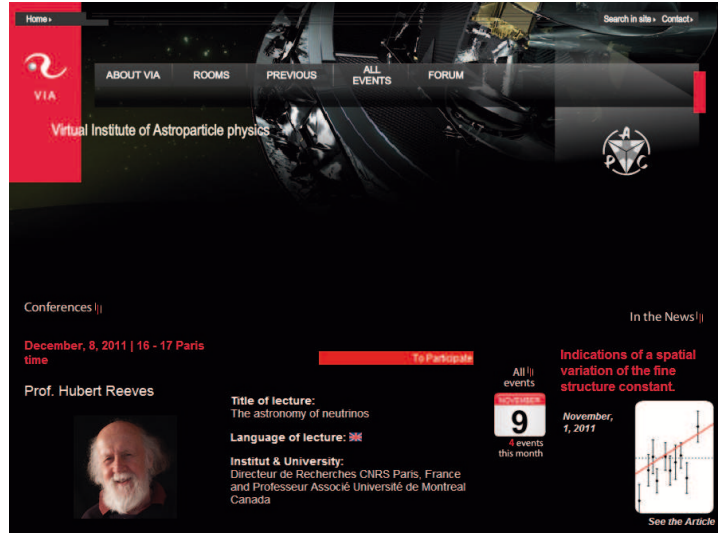}
        \caption{The home page of VIA site}
        \label{homevia}
    \end{center}
\end{figure}

The structure of
VIA complex is illustrated on Fig. \ref{homevia}.
The home page, presented on this figure, contains the information on
VIA activity and menu, linking to directories (along the upper line
from left to right): with general information on VIA (About VIA),
entrance to VIA virtual lecture hall and meeting rooms (Rooms), the
library of records and presentations (Previous) of VIA Lectures (Previous $\rightarrow$ Lectures), records of
online transmissions of Conferences(Previous $\rightarrow$ Conferences), APC Seminars (Previous $\rightarrow$ APC Seminars) and APC Colloquiums (Previous $\rightarrow$ APC Colloquiums)
and courses, Calender of the past and future VIA events
(All events) and VIA Forum (Forum). In the upper right angle there are links to Google search
engine (Search in site) and to contact information (Contacts). The
announcement of the next VIA lecture and VIA online transmission of
APC Colloquium occupy the main part of the homepage with the record
of the most recent VIA events below. In the announced time of the event
(VIA lecture or transmitted APC Colloquium) it is sufficient to
click on "to participate" on the announcement and to Enter as Guest
in the corresponding Virtual room. The Calender links to the program
of future VIA lectures and events. The right column on the VIA
homepage lists the announcements of the regularly up-dated hot news of Astroparticle
physics.

In 2010 special COSMOVIA tours were undertaken in Switzerland
(Geneva), Belgium (Brussels, Liege) and Italy (Turin, Pisa, Bari,
Lecce) in order to test stability of VIA online transmissions from
different parts of Europe. Positive results of these tests have
proved the stability of VIA system and stimulated this practice at
XIII Bled Workshop. These tours assumed special equipment, including, in particular, the use of the sensitive audio system KONFTEL 300W \cite{konftel}. The records of the videoconferences at the previous XIII Bled Workshop are available on VIA site \cite{VIAbled10}.

In 2011 VIA facility was effectively used for the tasks of the Paris Center of Cosmological Physics (chaired by G. Smoot), for the public programme "The two infinities" (conveyed by J.L.Robert) for Post-graduate programme assumed by the agreement between the University Paris Diderot and the University of Geneva. It has effectively supported participation at distance at meetings of the Double Chooz collaboration: the experimentalists, being at shift, took part in the collaboration meeting in such a virtual way.

It is assumed that the VIA Forum can continue and extend the
discussion of questions that were put in the interactive VIA events.
The Forum is intended to cover the topics: beyond the standard
model, astroparticle physics, cosmology, gravitational wave
experiments, astrophysics, neutrinos. Presently activated in
English, French and Russian with trivial extension to other
languages, the Forum represents a first step on the way to
multi-lingual character of VIA complex and its activity.

One of the interesting forms of Forum activity is the educational
work. For the last four years M.Khlopov's course "Introduction to cosmoparticle physics" is given in the form of VIA videoconferences and the records of these lectures and their ppt presentations are put in the corresponding directory of the Forum \cite{VIAforum}. Having attended the VIA course of lectures in order to be
admitted to exam students should put on Forum a post with their
small thesis. Professor's comments and proposed corrections are put
in a Post reply so that students should continuously present on
Forum improved versions of work until it is accepted as
satisfactory. Then they are admitted to pass their exam. The record
of videoconference with their oral exam is also put in the
corresponding directory of Forum. Such procedure provides completely
transparent way of estimation of students' knowledge.

\subsection{VIA lectures, online transmissions and virtual meetings} 
\begin{figure}
    \begin{center}
        \includegraphics[scale=0.3]{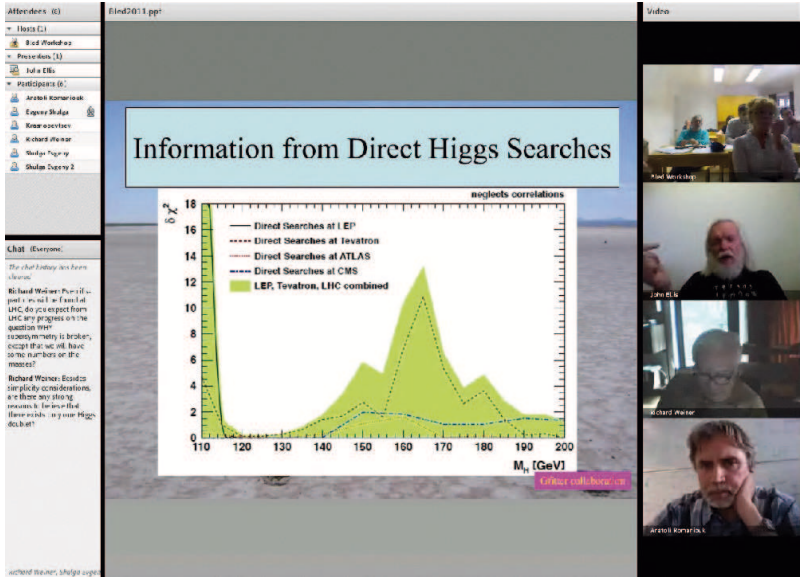}
        \caption{Videoconference Bled-Marburg-Liege-Geneva-Moscow-Paris with lecture by John Ellis,
        which he gave from his office in CERN,
        Switzerland, became a part of the program of
           XIV Bled Workshop.}
        \label{ellis}
    \end{center}
\end{figure}

First tests of VIA
system, described in \cite{Khlopov:2008vd,archiVIA,BledVIA9,BledVIA10},
involved various systems of videoconferencing. They included skype,
VRVS, EVO, WEBEX, marratech and adobe Connect. In the result of
these tests the adobe Connect system was chosen and properly
acquired. Its advantages are: relatively easy use for participants,
a possibility to make presentation in a video contact between
presenter and audience, a possibility to make high quality records
and edit them, removing from records occasional and rather rare
disturbances of sound or connection, to use a whiteboard facility
for discussions, the option to open desktop and to work online with
texts in any format. The regular form of VIA meetings assumes that
their time and Virtual room are announced in advance. Since the
access to the Virtual room is strictly controlled by administration,
the invited participants should enter the Room as Guests, typing
their names, and their entrance and successive ability to use video
and audio system is authorized by the Host of the meeting. The
format of VIA lectures and discussions is shown on Fig. \ref{ellis},
illustrating the talk "New physics and its experimental probes" given by John Ellis from CERN in the framework
of XIV Workshop. The complete record of this talk and other VIA
discussions are available on VIA website \cite{VIAbled11}.

The ppt or pdf file of presentation is uploaded in the system in
advance and then demonstrated in the central window. Video images of
presenter and participants appear in the right window, while in the
upper left window the list of all the attendees is given. To protect
the quality of sound and record, the participants are required to
switch out their microphones during presentation and to use lower
left Chat window for immediate comments and urgent questions. The
Chat window can be also used by participants, having no microphone,
 for questions and comments during Discussion. The interactive form of VIA lectures provides oral discussion, comments and questions during the lecture. Participant should use in this case a "raise hand" option, so that presenter gets signal to switch our his microphone and let the participant to speak. In the end of presentation
 the central window can be used for a whiteboard utility
 as well as the whole structure of windows can be changed,
 e.g. by making full screen the window with the images of participants of discussion.

 Regular activity of VIA as a part of APC includes online transmissions of all the APC Colloquiums and of some topical APC Seminars, which may be of interest for a wide audience. Online transmissions are arranged in the manner, most convenient for presenters, prepared to give their talk in the conference room in a normal way, projecting slides from their laptop on the screen. Having uploaded in advance these slides in the VIA system, VIA operator, sitting in the conference room, changes them following presenter, directing simultaneously webcam on the presenter and the audience.

\section{\label{Bled} VIA Sessions at Bled Workshop}
\subsection{The program of discussions}
In the course of XIV Bled Workshop meeting the list of open questions
was stipulated, which was proposed for wide discussion with the use
of VIA facility.

The list of these questions was put on VIA Forum (see \cite{VIAforum11}) and all the
participants of VIA sessions were invited to address them during VIA
discussions. Some of them were covered in the VIA lecture "New physics and its experimental probes" given by John Ellis (see the records in \cite{VIAbled11}). During the XIV Bled Workshop the test of minimal necessary equipment was undertaken. VIA Sessions were supported by personal laptop with WiFi Internet connection only.
It proved the possibility to provide effective interactive online VIA videoconferences even in the absence of any special equipment. Only laptop with microphone and webcam together with WiFi Internet connection was shown to be sufficient not only for attendance, but also for VIA presentations and discussions.

Another application at Bled Workshop was related with VIA records of closed meetings. The presentation was given in the regime of VIA online transmission and recorded, but the admission to the virtual room was restricted by a very short list of distant participants and the link to the record was available to a restricted list of users. Such use of VIA facility may be of interest for closed collaboration meetings.

\subsection{VIA discussions}
VIA sessions of XIV Bled Workshop have developed from
the first experience at XI Bled Workshop \cite{Bregar:2008zz} and
their more regular practice at XII and XIII Bled Workshops \cite{BledVIA9,BledVIA10}.
They became a regular part of the Bled Workshop's programme.

\begin{figure}
    \begin{center}
        \includegraphics[scale=0.3]{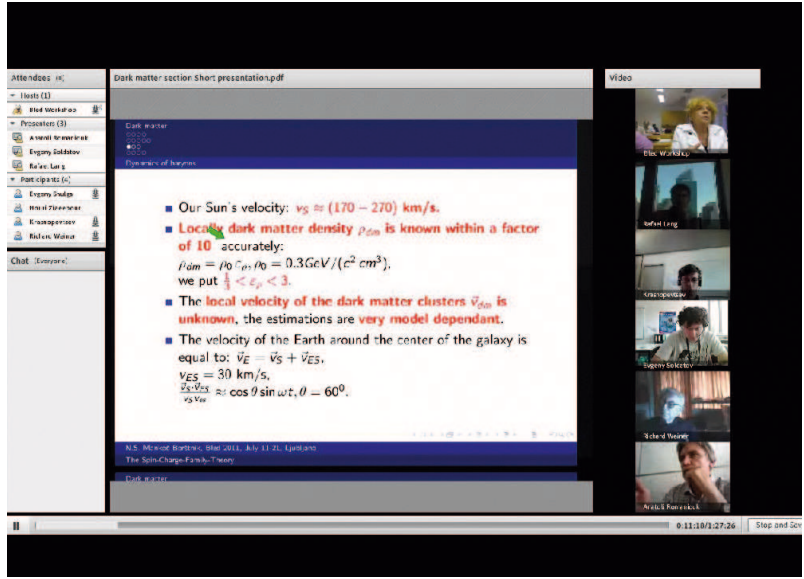}
        \caption{Bled Conference Discussion Bled-Moscow-CERN-UK-Marburg-Liege-USA}
        \label{dm}
    \end{center}
\end{figure}

 In the framework of the program of XIV Bled Workshop, John Ellis, staying in his office in CERN,
 gave his talk "New physics and its experimental probes" and took part in the
 discussion, which provided a brilliant demonstration of the interactivity of VIA in the way
most natural for the non-formal atmosphere of Bled Workshops. The
advantage of the VIA facility has provided distant
participants to share this atmosphere and contribute the discussion. VIA sessions were finished by the discussion of puzzles
 of dark matter searches (see \cite{VIAbled11}). N.S. Manko\v c Bor\v stnik and G. Bregar presented possible dark matter candidates
 that follow from the approach, unifying spins and charges, and Maxim Khlopov presented composite
 dark matter scenario, mentioning that it can offer the solution for the puzzles of direct
 dark matter searches as well as that it can find physical basis in the above approach. H.B.Nielsen informed about his macroscopic candidate for dark matter.
 The comments by Rafael Lang from his office in USA were very important
 for clarifying the current status of  experimental constraints on the possible properties
 of dark matter candidates (Fig. \ref{dm})

VIA sessions provided participation at distance in Bled discussions
for John Ellis and A.Romaniouk (CERN, Switzerland), K.Belotsky, N.Chasnikov,
A.Mayorov and E. Soldatov (MEPhI, Moscow), J.-R. Cudell (Liege,
Belgium), R.Weiner (Marburg, Germany) H.Ziaeepour (UK), R.Lang (USA)  and many
others.

\section{Conclusions}

Current VIA activity is integrated in the structure of APC
laboratory and includes regular weekly videoconferences with VIA
lectures, online transmissions of APC Colloquiums and Seminars, a
solid library of their records and presentations, together with the work of
multi-lingual VIA Internet forum.

The Scientific-Educational complex of Virtual Institute of
Astroparticle physics can provide regular communications between
different groups and scientists, working in different scientific
fields and parts of the world, get the first-hand information on the
newest scientific results, as well as to support various educational
programs at distance. This activity would easily allow finding
mutual interest and organizing task forces for different scientific
topics of astroparticle physics and related topics. It can help in
the elaboration of strategy of experimental particle, nuclear,
astrophysical and cosmological studies as well as in proper analysis
of experimental data. It can provide young talented people from all
over the world to get the highest level education, come in direct
interactive contact with the world known scientists and to find
their place in the fundamental research.
VIA applications can go far beyond the particular tasks of astroparticle physics and give rise to an interactive system of mass media communications.

VIA sessions became a natural part of a program of Bled Workshops,
opening the room of discussions of physics beyond the Standard Model
for distant participants from all the world. The experience of VIA applications at Bled Workshops plays important role in the development of VIA facility as an effective tool of $e-science$ and $e-learning$.
\section*{Acknowledgements}
 The initial step of creation of VIA was
 supported by ASPERA. I am grateful to P.Binetruy, J.Ellis and S.Katsanevas for
 permanent stimulating support, to J.C. Hamilton for support in VIA
 integration in the structure of APC laboratory,
to K.Belotsky, A.Kirillov and K.Shibaev for assistance in
educational VIA program, to A.Mayorov, A.Romaniouk and E.Soldatov for fruitful
collaboration, to M.Pohl, C. Kouvaris, J.-R.Cudell,
 C. Giunti, G. Cella, G. Fogli and F. DePaolis for cooperation in the tests of VIA online
 transmissions in Switzerland, Belgium and Italy and to D.Rouable for help in
 technical realization and support of VIA complex.
 I express my gratitude to N.S. Manko\v c Bor\v stnik, G.Bregar, D. Lukman and all
 Organizers of Bled Workshop for cooperation in the organization of VIA Sessions at XIV Bled Workshop.


\end{document}